\documentclass[a4paper,preprintnumbers,floatfix,superscriptaddress,pra,twocolumn,showpacs,notitlepage,longbibliography]{revtex4-1}
\usepackage[utf8]{inputenc}
\usepackage[T1]{fontenc}
\usepackage[sc,osf]{mathpazo}
\usepackage{mathtools}
\usepackage{amsfonts}
\usepackage{bbm} 
\usepackage{booktabs}
\usepackage{xcolor}
\usepackage{diagbox}
\usepackage{float}
\usepackage{lipsum}

\begin{document}

\title{Statistical properties and repetition rates for a quantum network with geographical distribution of nodes}
\author{Rute Oliveira}
\affiliation{International Institute of Physics, Federal University of Rio Grande do Norte, 59078-970, Natal, Brazil}
\author{Raabe Oliveira}
\affiliation{Departamento de Física, Universidade Federal do Ceará, 60451–970, Fortaleza, Ceará, Brazil}
\author{Nadja K. Bernardes}
\affiliation{Departamento de Física, Universidade Federal de Pernambuco, Recife, Brazil}
\author{Rafael Chaves}
\affiliation{International Institute of Physics, Federal University of Rio Grande do Norte, 59078-970, Natal, Brazil}
\affiliation{School of Science and Technology, Federal University of Rio Grande do Norte, Natal, Brazil}
\date{\today}

\begin{abstract}
Steady technological advances and recent milestones such as intercontinental quantum communication and the first implementation of medium-scale quantum networks are paving the way for the establishment of the quantum internet, a network of nodes interconnected by quantum channels. Here we build upon recent models for quantum networks based on optical fibers by considering the effect of a non-uniform distribution of nodes, more specifically based on the demographic data of the federal states in Brazil. We not only compute the statistical properties of this more realistic network, comparing its features with previous models but also employ it to compute the repetition rates for entanglement swapping, an essential protocol for quantum communication based on quantum repeaters. 
\end{abstract}

\maketitle

\section{Introduction}
By harnessing the principles of quantum mechanics, emerging quantum technologies have the potential to revolutionize various fields, from computing \cite{dalzell2023quantum} and communication \cite{gisin2007quantum} to sensing \cite{degen2017quantum} and simulation of physical systems \cite{georgescu2014quantum}. Quantum technologies exploit the unique quantum properties exhibited by the microscopic world, such as superposition and entanglement, to perform tasks that were previously thought to be impossible or highly impractical using classical methods. 

In particular, quantum communication within quantum networks has transitioned from theoretical concepts to practical applications, standing out among the most mature and well-developed quantum technologies. A quantum network consists of distant parties connected by quantum channels 
through which quantum bits, typically encoded in various degrees of freedom of photons \cite{gisin2007quantum}, can be exchanged. Optical fibers \cite{valivarthi2016quantum,wengerowsky2019entanglement} and satellite-to-ground \cite{Bedington2017,Yin1140,PhysRevLett.120.030501,sidhu2021advances} are the most promising candidates for establishing quantum channels, since breakthrough experimental advances now allow for quantum communication and the sharing of quantum entanglement through large distances, paving the way for undergoing development of the quantum internet \cite{kimble2008quantum,simon2017towards,wehner2018quantum,caleffi2018quantum}. The advantages of such quantum networks are illustrated by several successful experimental implementations of quantum key distribution (QKD) \cite{Gisin_2002}, distributed computing \cite{buhrman2003distributed,ho2022entanglement}, Bell inequalities violations \cite{hensen2015loophole,loop1,loop2} and quantum teleportation \cite{pirandola2015advances}, also including applications such are clock synchronization \cite{ClockSinc} and
private quantum computation on a cloud \cite{Broadbent_2009,PhysRevA.96.012303}.

In the face of all advances, several initiatives for the development of quantum networks of growing size and complexity have been established around the world such as the \emph{European Quantum Internet Alliance} (see Ref. \cite{riedel2018europe}) and \emph{America’s Blueprint for the Quantum Internet} (see Ref. \cite{awschalom2020long}). Of particular relevance is the \emph{Quantum Experiments at Space Scale} \cite{lu2022micius}, a Chinese research project that with the launch of the Micius satellite was able to distribute entangled photons over record distances \cite{Yin1140,ren2017ground} and establish the first integrated space-to-ground quantum communication network connecting 4 different cities and spanning distances over 4,600 kilometers \cite{chen2021integrated}.

With the ongoing development of the quantum Internet, it becomes essential to understand the properties such a new and unexplored kind of network will have \cite{brito2020statistical,brito2021satellite,zhuang2021quantum,azuma2021tools,harney2022analytical,bugalho2023distributing,wei2022towards,nokkala2020probing,zhang2021quantum}. A task for which network science offers a natural and powerful set of tools \cite{albert2002statistical,barabasi2016network,nokkala2023complex}. For instance, the connectivity of the network tells us whether it is possible to transmit information across the whole network while the average distance between nodes informs how efficiently communication can be achieved. Within this context, two different kinds of photonic networks have been analyzed. Considering an optical-fiber-based network \cite{brito2020statistical}, it was shown that even a very small density of nodes is sufficient to produce fully connected photonic networks. However, the typical distances between nodes increase in a power-law relation with the number of nodes, meaning that it does not lead to the small-world property, an undesired property since such large network distances imply that more entanglement swappings \cite{zukowski1993event,pan1998experimental} and consequently more quantum repeaters \cite{briegel1998quantum,sangouard2011quantum,ruihong2019research,Azuma22} are needed if one wants to distribute entanglement among nodes in the network. Nicely, however, if the links interconnecting the different nodes are quantum channels mediated by a satellite, such as the Micius \cite{lu2022micius}, the corresponding quantum network display hubs, nodes that have a large number of connections and have the effect of decreasing the network distances and making the network more robust against node and link random failures \cite{brito2021satellite}, a clear advantage in entanglement distribution.

In those studies, however, the models assumed a uniform distribution of nodes across the whole area of the network. In practice, due to the geographical and demographical properties of the regions where quantum networks are to be established, a non-uniform distribution of nodes is to be expected. Another crucial aspect is to move beyond the statistical properties and understand the repetition rates allowed by quantum repeaters in the network. These are the two new ingredients we address in this work considering an optical-fiber-based network. First, we analyze the statistical effects of a non-uniform distribution of nodes based on the demographical and geographical properties of Brazil \cite{ipea}. Second, using the analytical analysis in \cite{bernardes2011rate}, we compute the repetition rates that can be achieved in this more realistic model of a quantum network.

The paper is organized as follows. In Sec. \ref{sec:sec2} we describe the construction of a quantum network with a non-uniform distribution of nodes, comparing its main network properties with those of a uniformly distributed network \cite{brito2020statistical}. In Sec. \ref{sec:sec3} we compute the repetition rates for quantum repeaters allowed by the uni-
form and non-uniform networks. In Sec. \ref{sec:sec4} we discuss our main findings also pointing out relevant directions for future research.

\begin{figure*}[t!]
\begin{center}
\includegraphics[scale=.85]{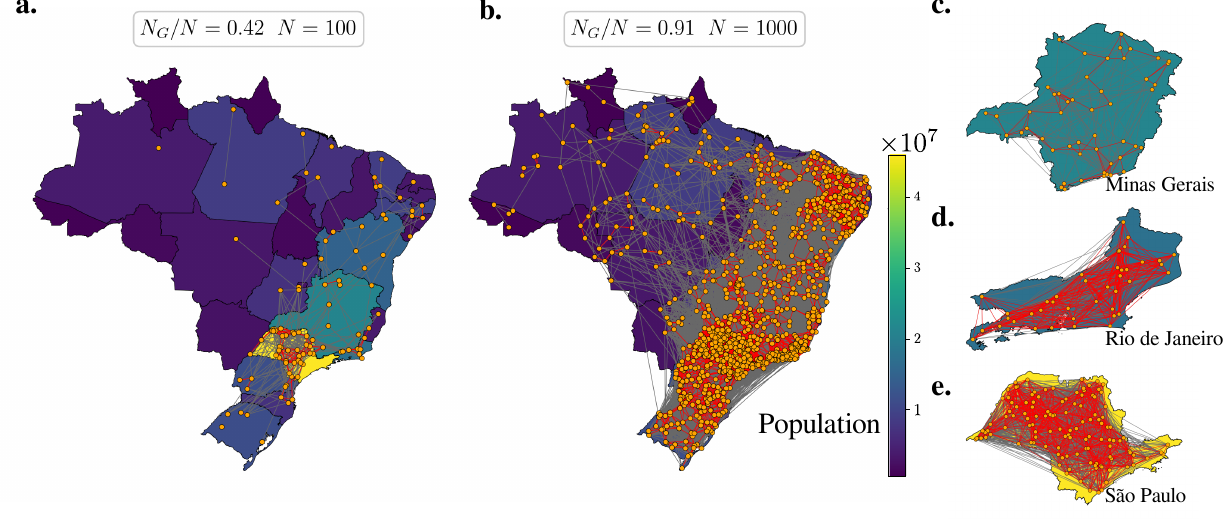}
\end{center}
\caption{ {\bf Samples of the non-uniform node distribution quantum network.} Grey edges represent the fiber-optics network generated in the third step, employing the Waxman model to distribute the edges in the whole Brazilian area. Red edges display the photonic links established in the fourth step, where we simulate the transmission of photons through it. $N_G$ refers to the number of nodes belonging to the biggest cluster in the network, and $N$ is the total number of nodes. As we progressively increase the size of the network (the number of nodes, since the area over which they are distributed is fixed), the emergence of the giant cluster becomes evident. The plots considered Brazil's territorial area as $A = 8.516.000$ km$^2$, wherein each node was inserted considering the population density of each Brazilian state, thus achieving a more realistic non-uniform distribution of nodes, \textbf{(a)} and \textbf{(b)} $N = 100$ and $N = 1000$ respectively, \textbf{(c)}, \textbf{(d)} and \textbf{(e)} $N = 500$ nodes for the three most populous Brazilian states.}
\label{fig:networkbrazil}
\end{figure*}

\section{Statistical properties of a quantum network}
\label{sec:sec2}

\subsection{Building a model for a quantum network}
A network model is characterized by a set of $N$ nodes (or sites) connected through edges (or links) obeying a given connection rule. Our first objective here is to reproduce the optical-fiber-based quantum network model introduced in Ref. \cite{brito2020statistical}. However, differently from there, we do not assume that the nodes are uniformly distributed and instead use geographic data and information to distribute the nodes according to the population density of Brazil. Brazil is an interesting case of study because it is a continental country with very different demographic densities depending on the region.

We consider a network built from optical fibers, which are one of the main candidates to carry quantum information encoded in photons. Our non-uniform model
(see Fig.~\ref{fig:networkbrazil})
is constructed through four stages: 

($1$) Using the package developed by a team at the Institute for Applied Economic Research (Ipea) \cite{ipea}, we approximate Brazil's area and map (and of each of its federative states) by polygons.

($2$) Using this approximated map, we distribute $N$ nodes over its whole area, ensuring their allocation follows the population density of the states in Brazil. However, within each state, the nodes are distributed uniformly.

($3$) Following \cite{lakhina2002geographic}, we simulate how the optical fibers are distributed among the nodes of the network using the Waxman model \cite{waxman1988routing}, which considers that each pair of nodes $i$ and $j$ are connected by a fiber (grey lines at Fig.~\ref{fig:networkbrazil}) with probability given by
\begin{equation}
\label{eq:waxprob}
    \prod_{ij} = \beta e^{-d_{ij}/\alpha L},
\end{equation}
where $0 < \beta \leq 1$ controls the average degree of the network. The parameter $\alpha > 0$ governs the characteristic edge length of the network, which corresponds to the maximum distance between any two directly connected nodes, $L$ is the maximum distance between any two nodes and $d_{ij}$ is the Euclidean distance, in kilometers, between site $i$ and the site $j$. That is, $d_{ij}$ represents the size of the optical fiber between stations $i$ and $j$. The values of constants $\alpha$ and $\beta$ have been estimated for specific optical fiber networks and in the present work we employ the values
estimated in \cite{lakhina2002geographic, durairajan2015intertubes} and given by $\alpha L = 226$ km and $\beta = 1$.

($4$) After generating the fiber-optics network, we proceed to simulate the transmission of photons through it, considering the unavoidable effects of photonic loss, which increase exponentially with the fiber length \cite{gisin2015far}. We incorporate this feature into the model by considering the probability that a photon is not lost, given by
\begin{equation}
    p_{ij} = 10^{-\gamma d_{ij}/10},
\end{equation}
where $d_{ij}$ is the Euclidean distance, in kilometers, between site $i$ and $j$ and $\gamma$ is the fiber loss (thus depending on the material and technology of the fiber). For instance, a silicon fiber has $\gamma \simeq 0.2$, the parameter we consider in our simulations. With that into account, the probability that a given pair of nodes linked by a fiber will also be connected through photons is given by
\begin{equation}
    P_{ij} = 1 - (1 - p_{ij})^{n_p}.
\end{equation}
In this expression, the number of photons transmitted between each node is regulated by the independent parameter $n_p$, that is, a photonic link is created between the nodes $i$ and $j$  if at least one out of $n_p$ photons is transmitted between them. We have employed $n_p = 1000$, since it ensures the establishment of connections spanning over $100$ km, the typical situation observed in practice. Apart from this practical consideration for the choice of $n_p$, we notice that previous research conducted in \cite{brito2020statistical, brito2021satellite} demonstrates that altering the values of $n_p$ does not change the qualitative properties of the model.

To obtain a diverse range of instances for this quantum network model, a substantial number of sample networks were analyzed (at least $10^3$ iterations). This extensive analysis ensured the generation of numerous distinct instances, allowing us to calculate the relevant properties of the network in a statistically meaningful manner.

\begin{figure}[t!]
\begin{center}
\includegraphics[scale=.46]{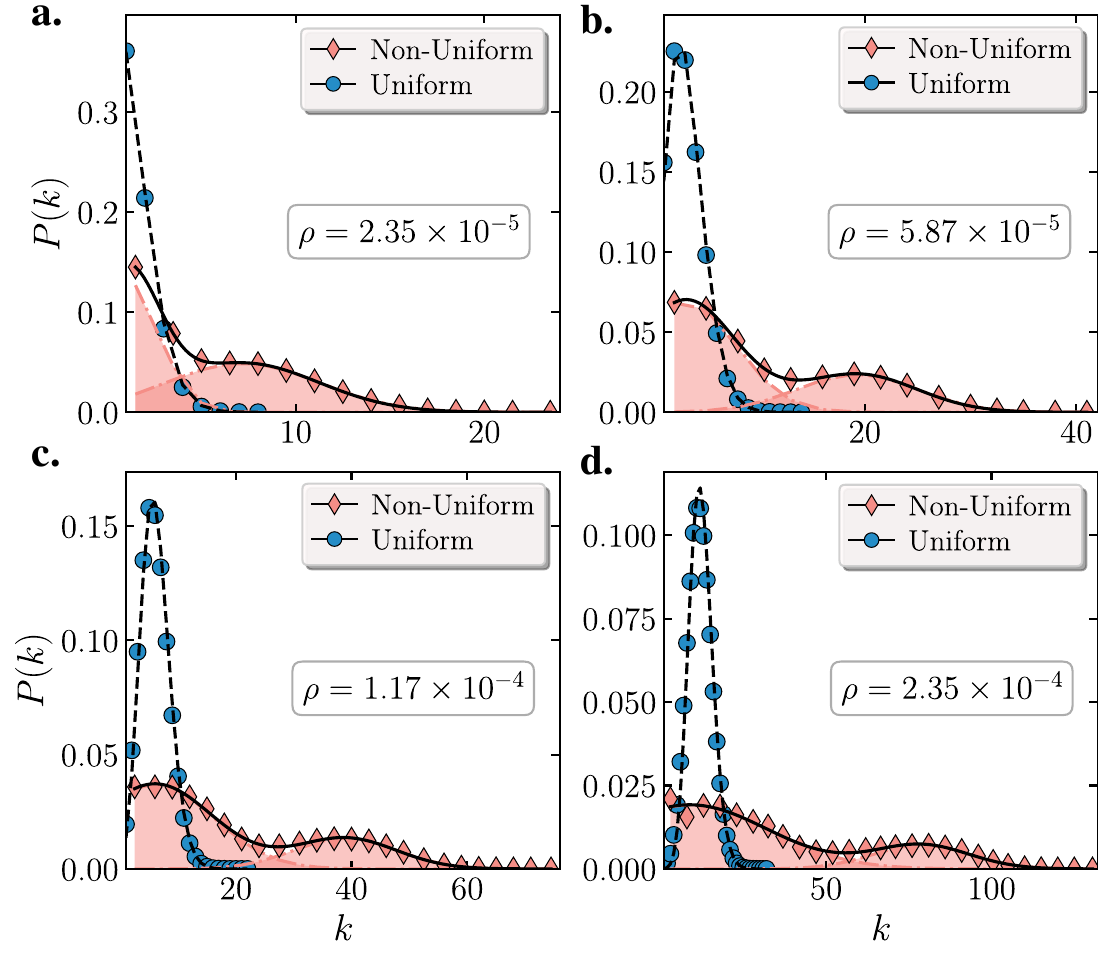}
\end{center}
\caption{ {\bf Comparison for the degree distribution $P(k)$ between the uniform and non-uniform distribution of nodes}. Several values of $N$, and consequently $\rho$ are employed: $(a)$ $\rho = 2.35 \times 10^{-5}$, $(b)$ $\rho = 5.87 \times 10^{-5}$, $(c)$ $\rho = 1.17 \times 10^{-4}$ and $(d)$ $\rho = 2.35 \times 10^{-4}$. For the uniform distribution network, $P(k)$ is represented by the blue circles, following a Poissonian. For the non-uniform case, we observed the presence of multiple peaks in the degree distribution curves, where we adjusted the points using a two-Gaussian fitting. The degree distribution was rescaled used linear bins ranging from $16$ to $28$ bins according to the network size $N$.}
\label{fig:fig2}
\end{figure}

\begin{figure}[t!]
\begin{center}
\includegraphics[scale=.46]{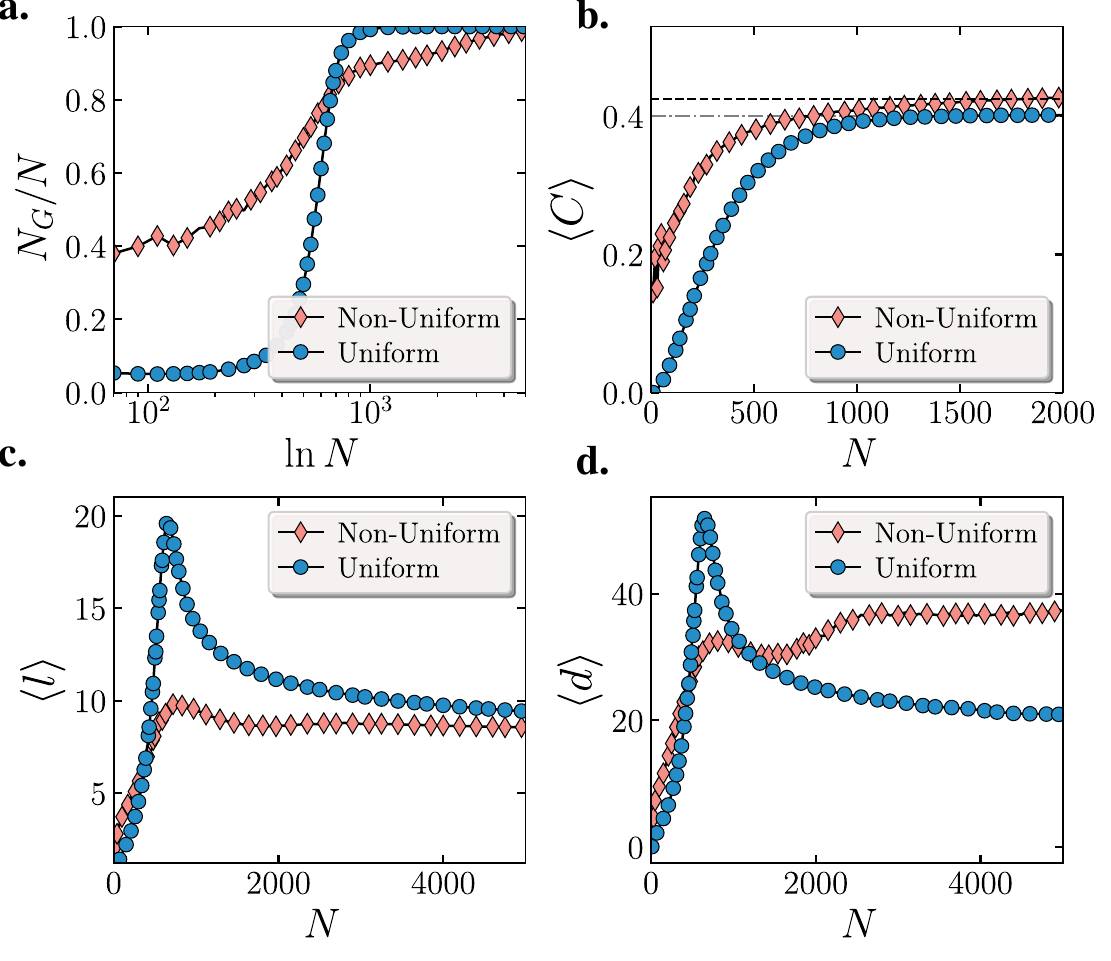}
\end{center}
\caption{ {\bf Comparison of relevant network properties for the uniform and non-uniform models} $(a)$ Relative size of the giant cluster as a function of $\ln{N}$. The uniform distribution network shows a sharp second-order transition from the disconnected to the connected phase, a feature not displayed in the non-uniform case. $(b)$ A similar behavior is obtained for both models regarding the average clustering coefficient $ \langle C \rangle$. $(c)$ Average shortest path $\langle l \rangle$ as a function of $N$. In both models, we have an initial steep increase in the shortest path that then starts to slowly decrease. $(d)$ Average diameter $\langle d \rangle$. Interestingly, in both cases, the diameter initially increases. After reaching a peak, the diameter of the uniform network slowly decreases. In turn, the non-uniform model reaches a plateau before having a slow increase in its diameter. In all the figures, red diamonds and blue circles refer to the non-uniform and uniform distribution networks.}
\label{fig:fig3}
\end{figure}

\subsection{Comparison between non-uniform and uniform distribution of nodes}

In the following, we describe the main statistical properties of this model of a quantum network comparing its features with those of the first model for an optical fiber-based quantum internet \cite{brito2020statistical} where the distribution of nodes was considered uniform, corresponding to modification in the steps 1 and 2 of our algorithm for the construction of the network. More precisely, in \cite{brito2020statistical}  the network nodes were distributed uniformly in a circle of a given radius $R$ and area $A$. In our comparison, we consider a radius $R = 1646.4$ Km in the original uniform distribution model, since it corresponds to an area that approximately corresponds to the geographical area of Brazil (considered in our non-uniform model) given by $A= 8.516.000$ km$^2$. 

The most commonly studied property in network science is its degree distribution, $P(k)$, which corresponds to the probability of finding a node with $k$ connections. Figure \ref{fig:fig2} gives the degree distribution for several values of $N$, and consequently $\rho=N/A$, the density of nodes per area. As proven in \cite{brito2020statistical}, the uniform model follows a Poissonian distribution given by
\begin{equation}
   P(k) = \frac{e^{-A\rho}(A\rho)^k}{k!},
\end{equation}
indicating that the majority of nodes will have a connectivity level close to $\langle k \rangle $, with deviations exponentially decreasing as the network size increases. In contrast, the connectivity distribution of the non-uniform distributed nodes in the optical-fiber-based quantum network exhibits a two-peak behavior, where the nodes have connectivity close to $\langle k_1 \rangle $ or $\langle k_2 \rangle $. As shown in Fig. \ref{fig:fig2}, the degree distribution can be well-fitted using a two-Gaussian distribution, given by
\begin{align}
P(k) = A_1   \left(\frac{1}{{\sigma_1   \sqrt{2\pi}}}\right)   \exp\left[-\frac{1}{2}\left(\frac{{x - \mu_1}}{{\sigma_1}}\right)^2\right] \notag \\ 
+ A_2   \left(\frac{1}{{\sigma_2   \sqrt{2\pi}}}\right)   \exp\left[-\frac{1}{2}\left(\frac{{x - \mu_2}}{{\sigma_2}}\right)^2\right].
\end{align}

These multiple peaks in the degree distribution can be attributed to distinct geographical groups of nodes within the network, each characterized by its own average degree. These groups are more densely connected among themselves compared to the rest of the network. This phenomenon arises due to the inherent heterogeneity of the node positions present within the network, representing the varying degrees of connectivity among different subsets of nodes. We notice that this phenomenon, networks with multiple average values in their degree distribution, can be observed in various real-world examples as financial networks with dynamic thresholds \cite{qiu2010financial} and complex networks from pseudo-periodic time series \cite{valente2004two}. 


Figure \ref{fig:fig3}(a) shows the evolution of the relative size of the largest connected component as a function of density, where $\rho = N/A$. In this case, we consider the fixed area of Brazil as $A = 8.516.000$ km$^2$. Interestingly, when starting with a small network, the largest cluster in the non-uniform model is larger than the largest cluster in the uniform model. This is likely due to the non-uniform distribution of sites, as the majority of them are concentrated in the Southeast region of the country, resulting in shorter distances between each site compared to the uniform distribution model, where sites are randomly distributed on a disk. Therefore, in small uniform distribution networks, its nodes are farther apart as compared to the non-uniform case, reducing the probability of connection between stations. This occurs because, as shown in equation \eqref{eq:waxprob}, the success probability depends on the distance between the sites. Additionally, as we can observe in Figure \ref{fig:fig3}(a) and proven in \cite{brito2020statistical}, a second-order phase transition occurs for the uniform model, as new sites are added to the network. In the non-uniform case, however, there is no sharp or apparent transition from a disconnected to a fully connected network.

In addition, we also investigate the average clustering coefficient,  a property measuring how the neighbors of each node are connected between them on average, given by
\begin{equation}
    \langle C \rangle = \frac{1}{N} \sum_{i = 1}^N \frac{2 n _i}{k_i (k_i - 1)},
\end{equation}
where $n_i$ is the number of edges between the $k_i$ neighbors of the site $i$, $k_i (k_i - 1)$ is total possible number of edges between them and $N$ is the network size. The average clustering coefficient of the non-uniformly and uniformly distributed nodes displays quite similar behaviors with increasing with $N$, the non-uniform case reaching a slightly higher maximum value given by $\langle C \rangle \approx 0.425$ (see Fig. \ref{fig:fig3}$(b)$). This implies that this sort of quantum network exhibits a high level of clustering, indicating that its nodes are more likely to form clusters or tightly interconnected groups. 

Another important property of a network is its average shortest path, which represents the average distance between pairs of nodes. It can be calculated as
\begin{equation}
\langle l \rangle = \frac{2}{N(N - 1)} \sum_{i < j}d_{ij},
\end{equation}
where $d_{ij}$ denotes the minimum number of edges connecting vertices $i$ and $j$. In line with common practice in the literature, we compute the shortest path length only for nodes belonging to the giant (connected) cluster, once $d_{ij} = \infty$ for vertices outside the same cluster. As shown in Figure \ref{fig:fig3}$(c)$, in both the uniform and non-uniform cases the behaviors are qualitatively similar. The average shortest path initially increases with the number of nodes $N$ until it reaches a peak value and it starts to slowly decrease seemingly reaching a plateau. This shows that even though the network does not display the small-world property, the fact that we increase the density of nodes in a fixed area increases the interconnections between the nodes decreasing $\langle l \rangle$. As it turns out, the number of entanglement swappings required to interconnect any two nodes in the network is smaller in the non-uniform model with $\langle l \rangle \leq 10$ for all number of sites $N$.

\begin{figure}[t!]
\begin{center}
\includegraphics[scale=.6]{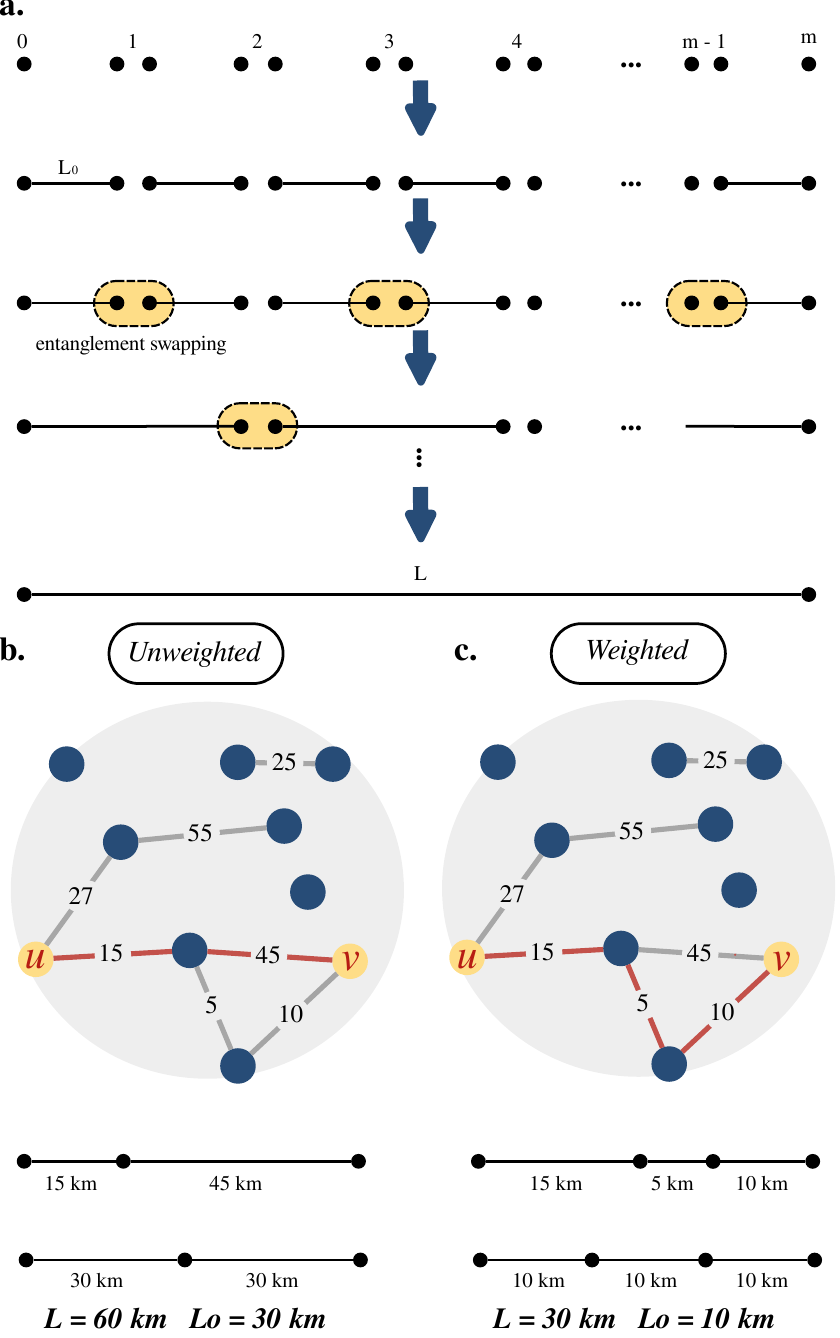}
\end{center}
\caption{ {\bf Idealized quantum repeater and choosing the minimum path.} 
\textbf{(a)} A total distance $L$ is divides in $2^n$ segments with length $L_0 = L/2^n$. Initially, entanglement is generated between neighboring repeater stations. The qubits at the intermediate stations are then connected. Finally, entanglement over the entire distance $L$ is obtained. \textbf{(b)} Exemplifies the total distance, denoted as $L$, under the assumption of unweighted connections, that is, we choose the path with the minimum number of intermediate steps. Consequently, the shortest distance between the vertices in yellow consists of two segments and $L=60$ km. \textbf{(c)} For the case of weighted edges between the nodes, we aim to minimize the Euclidian distance between the nodes. In the example, this corresponds to three segments and $L=30$ km.}
\label{fig:02}
\end{figure}

Finally, in network theory, the diameter represents the maximum distance between any pair of nodes, determined by the longest shortest path among all possible pairs. In the context of our study, this concept translates to the maximum number of entanglements between any two nodes in a quantum-connected network. While entanglement swapping allows for the potential entanglement of any two nodes through intermediate nodes, the presence of inevitable errors introduces noise to the entanglement during these intermediary processes. As a result, maintaining short distances in the network becomes crucial to preserve a higher level of entanglement between the desired end nodes, underscoring its significance. For example, as illustrated in Fig. \ref{fig:fig3}$(d)$, the diameter of the optical-fiber-based quantum network with non-uniformly distributed nodes is smaller in the peak region compared to the standard uniform distribution model. However, as the network size grows, the diameter gradually increases, likely a consequence of the fact that as we increase the number of nodes, distant geographical regions start to be connected thus increasing the diameter of the network.

\begin{figure*}[t!]
\begin{center}
\includegraphics[scale=.73]{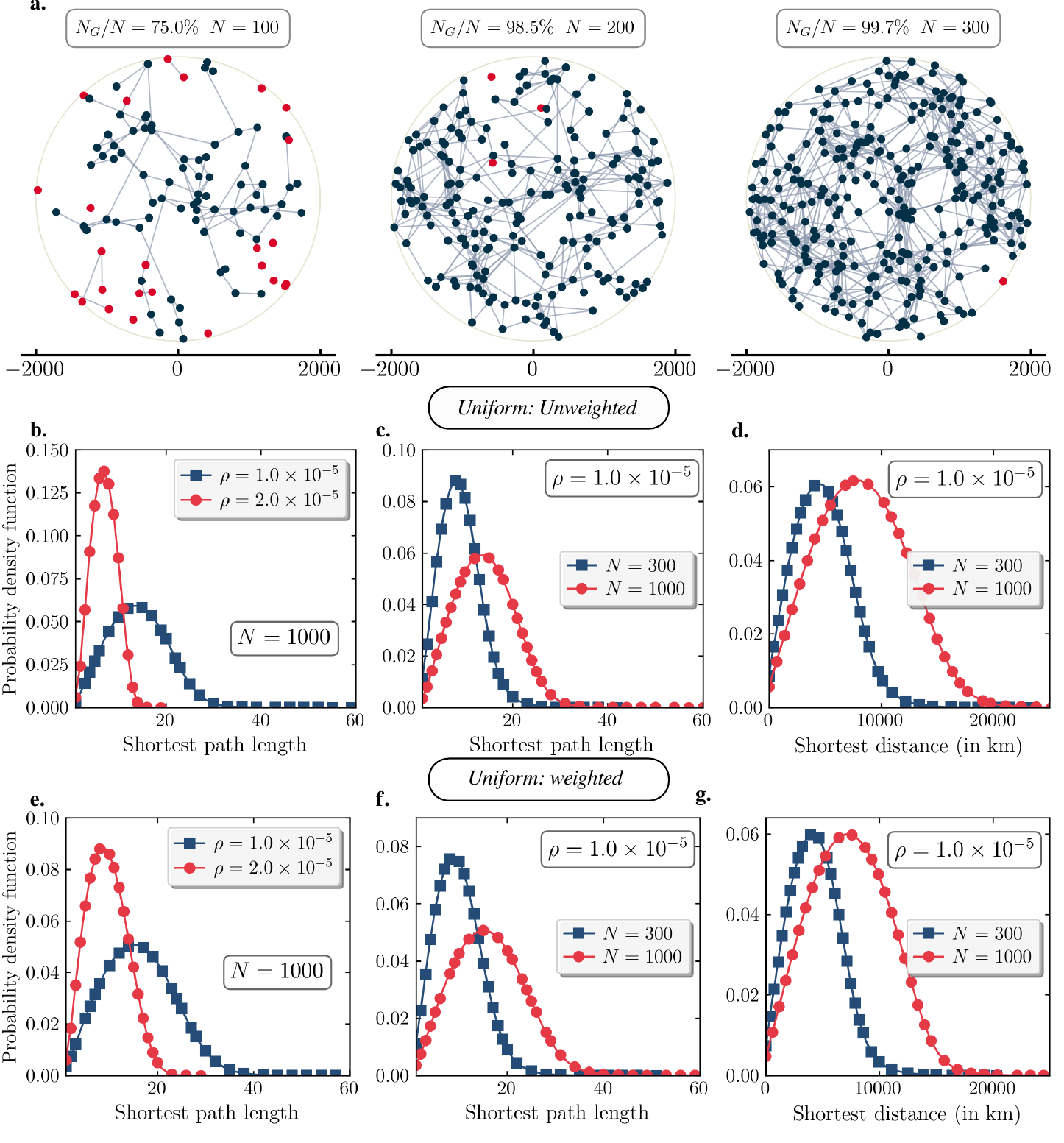}
\end{center}
\caption{ {\bf Samples from the Waxman model.} {\bf (a)} The gray links symbolize the fiber-optic channels and the blue vertices constitute the giant cluster. $N_G$ denotes the number of vertices belonging to the largest cluster in the network, and $N$ represents the total number of vertices. The figure takes into account a radius of $R = 1800$ km (approximately covering the area of the United States) and illustrates that the largest cluster comprises the entirety of the vertices when $N = 300$. {\bf (b)} Distance distributions for fixed $N = 1000$ some values of $\rho$ considering unweighted edges. {\bf (c)} Distance distributions for fixed $\rho = 1.0 \times 10^{-5}$ some values of $N$, also considering unweighted edges. {\bf (d)} Distance distributions for fixed $\rho = 1.0 \times 10^{-5}$ some values of $N$ for unweighted edges. {\bf (e)}, {\bf (f)} and {\bf (g)} show the same as in (b), (c) and (d) considering weighted edges.}
\label{fig:network}
\end{figure*}

\section{Repetition rates for optical fiber networks}
\label{sec:sec3}
Entanglement is the key feature in quantum information processing, in particular in communication protocols, since it enables to teleport quantum information \cite{bennett1993teleporting,pirandola2015advances}, communicate more efficiently \cite{buhrman2001quantum,ho2022entanglement} and securely \cite{scarani2009security}.  Unfortunately, in practice, the quantum channels over which the quantum systems are distributed are noisy, leading to an exponential decay of entanglement with the distance traveled by the photons. Typical classical procedures to revert the deleterious effects of noise, such as amplification or cloning, cannot be directly applied to quantum channels. To circumvent this issue and be able to perform long-distance quantum communication, quantum repeaters \cite{briegel1998quantum,sangouard2011quantum,ruihong2019research,Azuma22} were proposed. Rather than distributing entanglement over long distances, entanglement is generated in
smaller segments and a combination of entanglement swapping \cite{zukowski1993event} and entanglement purification \cite{pan2001entanglement} or quantum error correction \cite{PhysRevLett.112.250501} enables one to establish high-quality entangled pairs over long distances. Experimentally, quantum repeaters are still a challenge. Recently, a telecom-wavelength quantum repeater node based on a trapped-ion processor achieved a 50 km-long connection \cite{PhysRevLett.130.213601}.

\begin{figure*}[t!]
\begin{center}
\includegraphics[scale=.73]{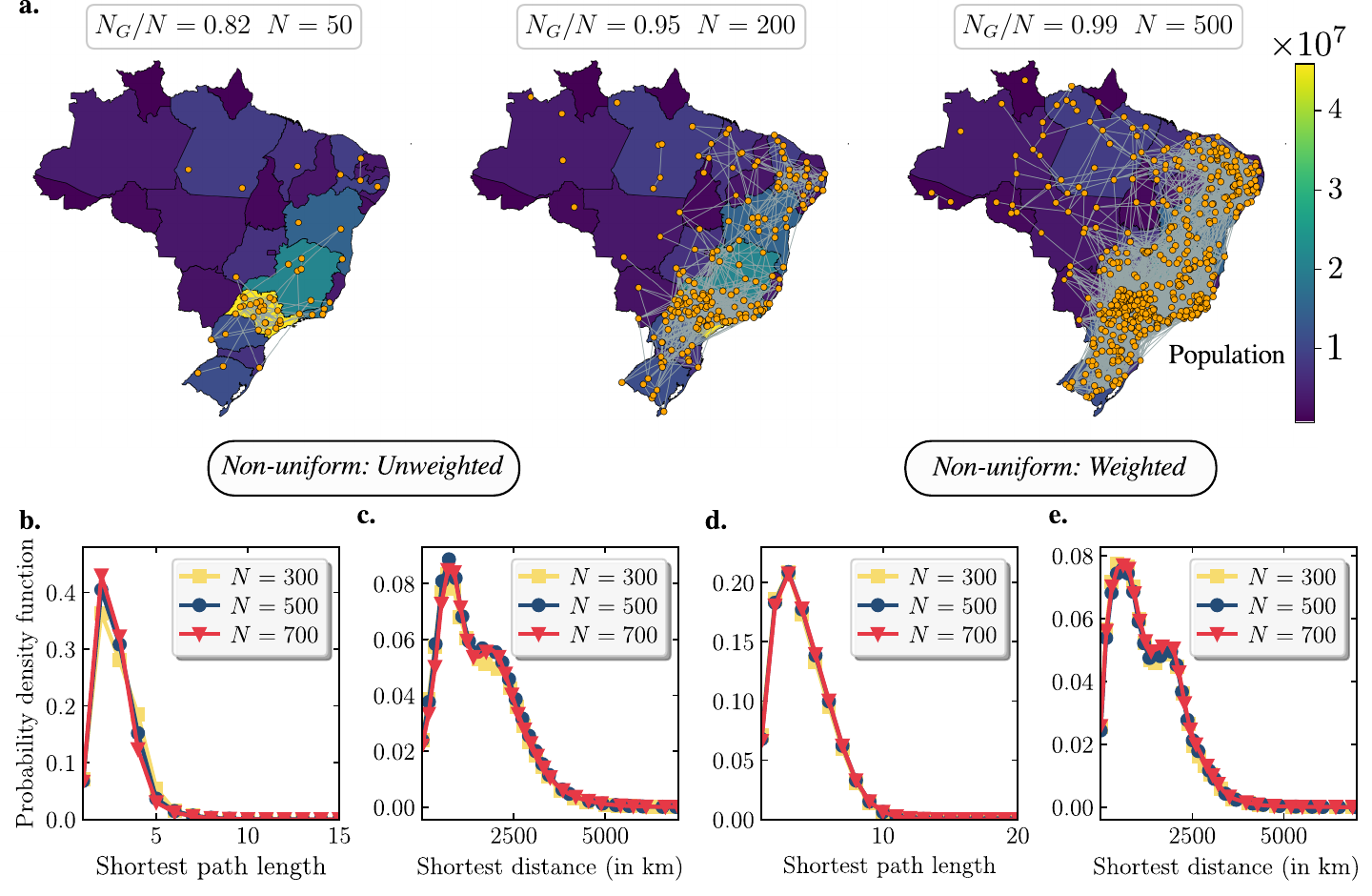}
\end{center}
\caption{\textbf{Samples of the non-uniform node distribution Waxman model}. \textbf{(a)} We distribute $N$ nodes over its whole area, ensuring their allocation follows the population density of the states in Brazil. However, within each state, the nodes are distributed uniformly. Edges represent the fiber-optics networks generated in the Eq.($1$). \textbf{(b)} Shortest Path Length for fixed $N = 300$, $N = 500$ and $N = 700$ considering unweighted edges for the non-uniform distribution of nodes. \textbf{(c)} Distance distributions for some defined network size values. \textbf{(d)} and \textbf{(e)} show the same as in (b) and (c) considering weighted edges.}
\label{fig:10}
\end{figure*}

\subsection{Computing the repetition rates between two nodes of the network}
 Within this context, one important figure of merit to evaluate the performance of the quantum repeater is the repetition rate that entanglement is generated \cite{bernardes2011rate,PhysRevA.87.052315,PhysRevA.100.032322,https://doi.org/10.1002/qute.201900141,liorni2021quantum,PhysRevA.105.012608, PhysRevA.107.012609}. Ref.~\cite{bernardes2011rate} derived analytical formulas for the time (and hence also for the repetition rates) needed to generate an entangled pair between two nodes at the opposite ends of a linear chain with a given number of intermediate nodes. In their analysis, it is assumed that optimal entanglement generation probabilities and deterministic
swapping routines are available, as well as
perfect memories, that is memories with infinite decoherence times. Employing the tools and results of reference \cite{bernardes2011rate}, our objective here is to compute the average repetition rate for our quantum network models.

As an illustrative example in Fig.\ref{fig:02}, a total distance $L$  is calculated between a pair of vertices ($u, v$) in the network, where exists $m$ segments separating the vertex $u$ to the vertex $v$. In this context, we employ an approximation, where each edge distance between ($u, v$) will be approximated by $L_0 = L/m$. First, entanglement is generated between the adjacent nodes, which is accomplished with probability $P_0$ (the initial probability at the output of a fiber link connecting nodes, $P_0=10^{-\gamma L_0/10}$ where $\gamma$ is the fiber loss which it depends on the photon wavelength). Then these segments are connected, extending the entanglement from $L_0$ to $2L_0$. This step is performed many times, until the terminal nodes, separated by $L = m L_0$, are entangled.

Given a pair of vertices ($u, v$), there will be typically many paths connecting them and we should choose the one minimizing the distance between the nodes. As for the distances between nodes in a network, there are two natural possibilities. First, consider the path with the smallest total length $L$, a weighted network (the weight being given by the Euclidian distance). Second, choose the path with the smallest number $m$ of intermediate segments, corresponding to $m-1$ intermediate nodes and unweighted connections since the distances between the intermediate nodes do not play a role in establishing this topological minimum distance. The difference between the two choices is illustrated in Fig.\ref{fig:02}. A priori, it is not clear which path is the optimal choice as we aim to maximize the repetition rate at which entangled pairs can be established between the nodes of the network. As the example in Fig.\ref{fig:02} shows, not necessarily the path with fewer segments -- thus requiring fewer entanglement swappings -- will be the one with the shortest physical Euclidian distance, the one minimizing the effects of photon loss.
 
To illustrate the general idea, let us calculate the rate for generating an entangled pair between two neighboring nodes in the network (the number of segments is m=1, corresponding to no intermediate node). If the distance between them is $L$ (in this case $L = L_0$ ), the average time necessary to generate an entangled pair is given by
\begin{equation}
    \langle T \rangle_1 = \frac{T_0}{P_0},
\end{equation}
where $T_0=2L_0 / c$ represents the least amount of time required to effectively establish entanglement across a distance $L_0$. This time encompasses the time for the photon for the entangled pair to cover the distance from one site to the other plus the time for the classical communication time essential for confirming the entanglement has been successfully implemented (for instance, the photon was not lost). The speed of light within an optical fiber is denoted by $c$ ($2 \times 10^8$ m/s).

For three connected vertices, $m = 2$, the average time necessary to generate an entangled pair at distance $L$ is then given by
\begin{equation}
    \langle T \rangle_2 = \frac{T_0}{P_0} \frac{(3 - 2P_0)}{(2 - P_0)}.
\end{equation}
Recall that the memories storing the photons are assumed to be ideal such that one successfully created pair can be kept until a second pair is created in the neighboring segment. Thus, for $m$ segments, the average time necessary to generate an entangled pair at distance $L$ is given by \cite{bernardes2011rate}
\begin{equation}
    \langle T \rangle_m = T_0 Z_m(P_0),
\end{equation}
where the average number of steps to successfully generate entanglement in all $m$ segments, $Z_m(P)$, is
\begin{equation}
    Z_m(P) = \sum_{j = 1}^{m}\binom{m}{j} \frac{(-1)^{j + 1}}{1 - (1 - P)^j},
\end{equation}
and $P$ is the probability of success given by
\begin{equation}
    P = 10^{-\gamma L_0 / 10},
\end{equation}
the probability entering the Waxman model, that is, of a photon traveling a distance $L_0$ without being lost. Finally, the rate at which we can successfully generate entanglement between the end nodes separated by a distance $L$ is given by
\begin{equation}
    R_m = \frac{1}{\langle T \rangle_m} = \frac{1}{T_0 Z_m(P_0)}.
\end{equation}

\subsection{Results}

We start considering the model of a quantum network introduced in \cite{brito2020statistical} with a uniform distribution of nodes. Examples of the network, considering a radius of $R=1800$ km and increasing the number of nodes, is shown in Fig. \ref{fig:network}(a). 

Figures \ref{fig:network}(b)-(g) refer to the distance distributions for the uniform distribution of nodes. In \ref{fig:network}(b)-(d), we present the distance distribution for the unweighted case, whereas in \ref{fig:network}(e)-(g), we depict these distributions considering the weighted scenario, where we employ the Dijkstra's algorithm \cite{dijkstra1959note} for distance calculations.

\begin{figure}
\begin{center}
\includegraphics[scale=.49]{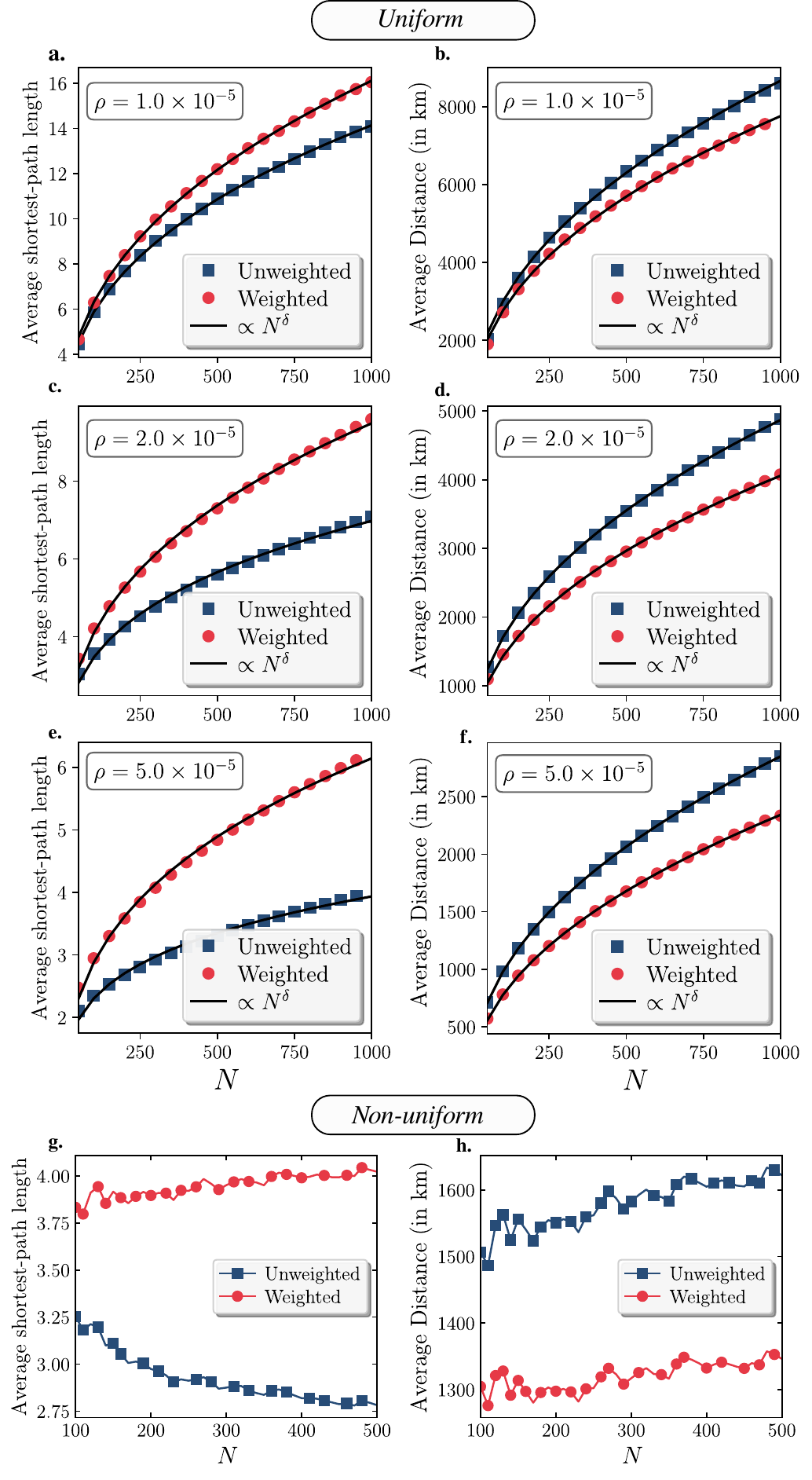}
\end{center}
\caption{\textbf{Average shortest path and average distances} \textbf{(a)} Average shortest-path length with $\rho = 1.0\times 10^{-5}$ for unweighted and weighted cases. \textbf{(b)} Average distance (in meters) with $\rho = 1.0\times 10^{-5}$ for both cases, unweighted and weighted. \textbf{(c)} same as show in (a) however we use $\rho = 2.0\times 10^{-5}$. \textbf{(d)} Average distance considering $\rho = 2.0\times 10^{-5}$. \textbf{(e)} and \textbf{(f)} Similarly to the previous ones, however, we increased the network density to $\rho = 5.0\times 10^{-5}$. In all uniform cases, we can describe these curves with a polynomial $N^{\delta}$ function (with $\delta$ shown in Table \ref{tab:1}). \textbf{(g)} Average shortest-path length considering the non-uniform distribution of nodes for both unweighted and weighted cases. \textbf{(h)} Average distance (in m) for the non-uniform distribution of nodes for both unweighted and weighted cases. In all cases, we observe that the average shortest-path length is lower when considering the weighted shortest path. However, this path is generally the one with the longest physical distance. As a result, the average distance becomes higher in the weighted case.}
\label{fig:caminho_medio}
\end{figure}

The Brazil network examples are shown in Figure \ref{fig:10}(a) with an area of approximately $8.516.000$ km$^2$ with an increase in the number of nodes, where we characterize this scenario for the non-uniform distribution of nodes. We show in Fig. \ref{fig:10}(b)-(e) the distributions of the shortest path length and shortest distance (in meters) considering unweighted and weighted edges with
a non-uniform distribution of nodes. We observe that for both the unweighted and weighted cases, there exists a characteristic value for the minimum number of links separating two vertices in the non-uniform network. In contrast, for physical distance, likewise to the degree distribution, we observe a two-peak behavior in the distribution. 

As illustrated in Figure \ref{fig:caminho_medio}, as expected, the average shortest-path length in the unweighted network is notably shorter compared to the average shortest-path length in the weighted scenario, for both the uniform and non-uniform cases. This finding aligns with our expectations, indicating that the absence of weight leads to more direct connections between nodes, because we do not assume the cost of that connection, resulting in shorter path lengths between all pairs of vertices of the giant cluster. Similarly, when we analyze the value of the least-cost distance, the weighted scenario exhibits a lower average distance compared to the unweighted case, because the cost of connections plays a pivotal role when we calculate the shortest path between nodes in a weighted network, where this result minimizes the physical distance on the path. Importantly, differently from the uniform distribution case, we see a clear difference between the weighted and unweighted choices in the non-uniform distribution of nodes.

Figures \ref{fig:caminho_medio}$(a)$, \ref{fig:caminho_medio}$(c)$ and \ref{fig:caminho_medio}$(e)$ depicts the average shortest path for the uniform distribution, showcasing different $\rho$ values. In contrast, Figure \ref{fig:caminho_medio}$(b)$, \ref{fig:caminho_medio}$(d)$ and \ref{fig:caminho_medio}$(f)$ shows the mean physical distance between all $N(N - 1)$ pairs of vertices in the giant cluster ($\langle l \rangle$), where $N$ is the number of sites in the giant cluster. Analogously, figure \ref{fig:caminho_medio}$(g)$ and \ref{fig:caminho_medio}$(h)$ show the case when we consider the non-uniform distribution of nodes. As can be seen, all curves for the uniform distribution are perfectly described by a polynomial $ N^{\delta}$. As expected, the case where we consider unweighted edges leads to a smaller average shortest path while weighted edges favor smaller average physical distances. This feature can also be seen in Fig.\ref{fig:network}(b)-(g) and Fig. \ref{fig:10}(b)-(e), displaying the average shortest path and distance distributions. In particular, for the case of the uniform distribution, the shortest path is described by a Poissonian with a larger variance in the unweighted case.

\begin{table}[]
\centering
\caption{The parameters of the polynomial $N^{\delta}$ using in the fit of the Fig.\ref{fig:caminho_medio}.} 
\label{tab:1}
\begin{tabular}{c|c|c|c|}
\cline{2-4}
                       & $\rho = 1.0 \times 10^{-5}$ & $\rho = 2.0 \times 10^{-5}$ & $\rho = 5.0 \times 10^{-5}$ \\ \hline
\multicolumn{1}{|c|}{Unweighted} & $\delta = 0.38$ & $\delta = 0.30$ & $\delta = 0.23$ \\ 
\multicolumn{1}{|c|}{Weighted}  & $\delta = 0.40$ & $\delta = 0.36$ & $\delta = 0.32$ \\ \hline
\end{tabular}
\end{table}

\begin{figure*}[t!]
\begin{center}
\includegraphics[scale=.5]{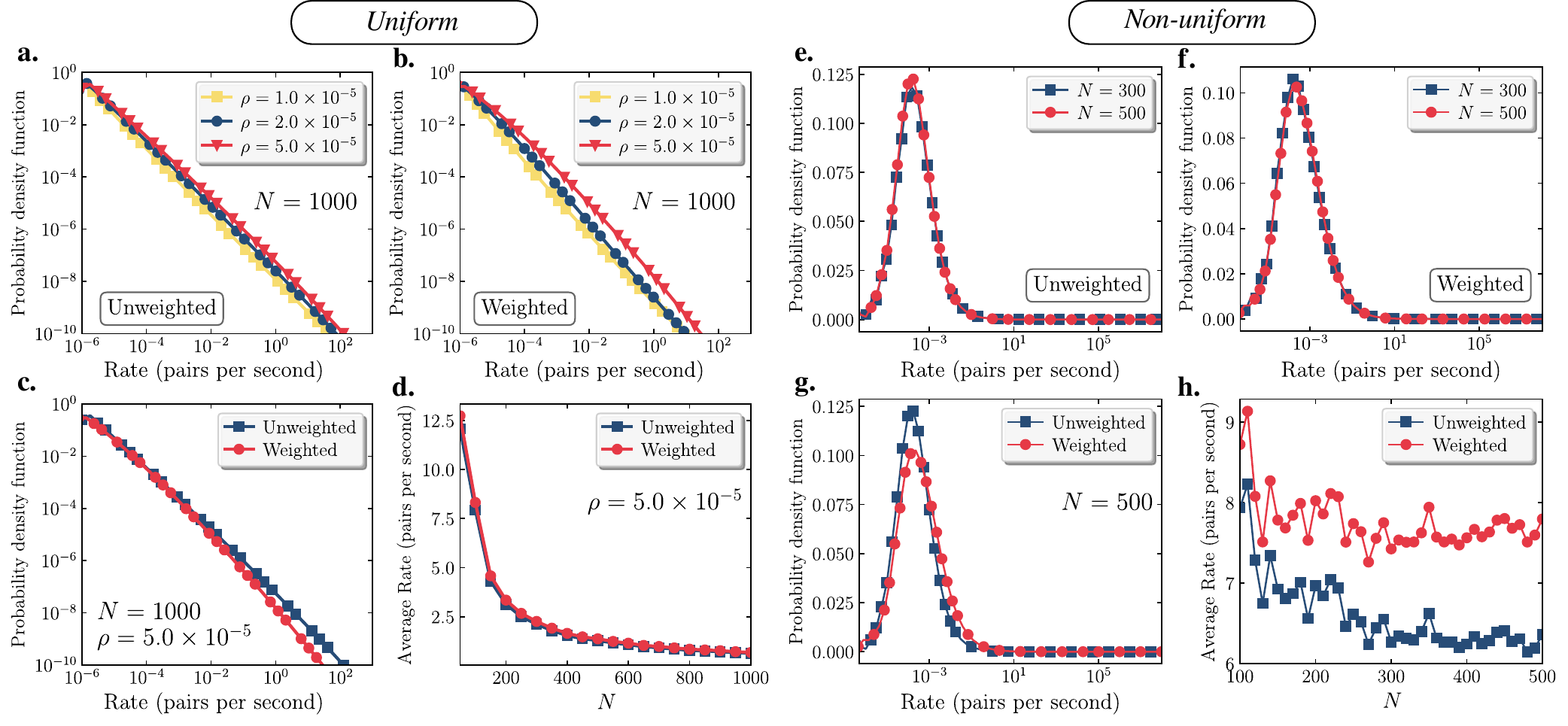}
\end{center}
\caption{\textbf{Repetition rates for a quantum network.} \textbf{(a)} Probability density function of the rates for fixed $N = 1000$ and typical values of $\rho = 1.0 \times 10^{-5}$, $2.0 \times 10^{-5}$ and $5.0 \times 10^{-5}$, where we do not consider the weights of the edges. \textbf{(b)} Same as in (a) however, the physical distance between nodes was taken into account (weighted case). \textbf{(c)} Probability density function of the rates for fixed $N = 1000$ and $\rho = 5.0 \times 10^{-5}$ for the both unweighted and weighted cases. \textbf{(d)} Average rate, $\overline{R_m}$, for the uniform case with $\rho = 5.0 \times 10^{-5}$ for the unweighted and weighted cases. \textbf{(e)} Rate distributions for fixed $N = 300$ and $N = 500$ considering unweighted edges, and \textbf{(f)} weighted edges. \textbf{(g)} Probability density function of the rates for fixed $N = 500$ considering the physical distance and not taking into account the distance between edges. \textbf{(h)} The average rate for unweighted and weighted edges for the non-uniform distribution.}
\label{fig:rates_values}
\end{figure*}

The average repetition rate to successfully generate entanglement between any two nodes in the network is given by
\begin{equation}
    \overline{R_m} = \frac{2}{n(n - 1)} \sum_{i < j} r_{ij},
\end{equation}
where $r_{ij}$ is the rate between the vertices $i$ and $j$. The average rate considering both the weighted and unweighted cases is shown in Fig. \ref{fig:rates_values}(d) for the uniform case and in Fig. \ref{fig:rates_values}(h) for the non-uniform case. In both cases, it is observed that weighted paths favor higher average repetition rates, a piece of relevant information for the optimization of such networks. As we can also see, the average repetition rate shows a rapid decay with increasing $N$ in the uniformly distributed network. For the non-uniform network, the decay of the rate with increasing $N$ is much less prominent. Interestingly, the probability density function for the rates is very different in the uniform and non-uniform networks. While in the uniform case, we see an exponential decay of the rate, in the non-uniform case the rate displays a Poissonian behavior, explaining the higher rates observed and the less prominent decay with an increasing number of nodes.

\section{Discussion}
\label{sec:sec4}
We have analyzed the properties of a quantum network based on the distribution of photonic quantum information via optical fibers. Differently from previous works \cite{brito2020statistical,brito2021satellite}, we consider the effect of a non-uniform distribution of network nodes, more specifically considering the demographic distribution of the federal states in Brazil. We not only consider its statistical properties but also employ the constructed network to analyze the repetition rates \cite{bernardes2011rate} that can be achieved for entanglement distribution between their nodes. We have assumed deterministic swapping and perfect memories. Note that in photonic schemes it is possible to achieve efficient swapping \cite{Azuma15, PhysRevA.92.052324}. Although perfect memories are an idealization, nitrogen-vacancy centers in diamonds are a good candidate for a memory qubit, because of the long coherence times of their nuclear spins. Indeed, they have been used for entanglement distillation between distant nodes \cite{doi:10.1126/science.aan0070} and to construct a multimode quantum network \cite{doi:10.1126/science.abg1919}.

Regarding its statistical properties, we observe that differently from the model with a uniform distribution of nodes, governed by a Poissonian distribution, the geographical network leads to a two-Gaussian degree distribution, those peaks being attributed to distinct groups of nodes within the network, each characterized by its average degree, an effect that is also observed in other real-world network examples \cite{qiu2010financial,valente2004two}. Nicely, the average shortest path and the diameter of the network are relatively small, achieving values of $\langle l \rangle \leq 10$ and $\langle d \rangle \leq 40$ even for a high concentration of network nodes ($N=2000$), meaning that on average few entanglement swappings are required to interconnect any two nodes in the whole network.

We have also compared the repetition rates that can be achieved in the uniform and non-uniform networks. For that, we first have to choose the path connecting two nodes. On one hand one could choose a weighted path, where the aim is to minimize the Euclidian distance between the nodes (minimizing the photon losses within the fibers). Alternatively, one can choose an unweighted connection, to minimize the number of edges between the nodes (and thus minimizing the number of required entanglement swapping to connect such nodes). In both the uniform and non-uniform networks, we observe that choosing weighted paths (minimizing the physical distances) leads to higher repetition rates, an effect that is more pronounced in the non-uniform network (a more realistic model of a real quantum network), providing a recipe for the actual implementation of protocols in future quantum networks. Interestingly, in the uniform network, the probability distribution of repetition rates shows an exponential decay while in the non-uniform case, the rate displays a Poissonian behavior, explaining the higher and more robust rates observed in the latter. The best average rate is $8$ pairs per second for the non-uniform and weighted case. Although this number is quite smaller compared to the all-photonic quantum repeater \cite{Azuma15} or one-way quantum repeater based on near-deterministic photon-emitter interfaces \cite{PhysRevX.10.021071}, where a rate of almost $70$ kHz is achievable for 1000 km, here this average rate is achieved considering a huge area ($8.516.000$ km$^2$) with reasonable amount of nodes (up to $N=500$). Furthermore, a few pairs of nodes in the network also reach rates in the kHz range.

Our results improve over previous studies \cite{brito2020statistical,brito2021satellite} by considering the effects of non-homogeneity in the nodes distribution and considering the repetition rates of entanglement swapping protocols, however, this should be seen as the first step towards more complicated and
realistic models. For instance, one could introduce the effects of non-perfect memories (finite coherence times), noise in the entanglement generation, or error in the swapping routines. More robust strategies such as quantum error correction \cite{PhysRevLett.112.250501} and multiplexing \cite{PhysRevLett.98.060502} could also be explored in this case. We hope our results motivate further work along these directions.

\section*{Acknowledgements} 
 This work was supported by the Serrapilheira Institute (Grant No. Serra-1708-15763), the Simons Foundation (Grant Number 1023171, RC), the Brazilian National Council for Scientific and Technological Development (CNPq) (INCT-IQ and Grant No 307295/2020-6)  and the Brazilian agencies MCTIC, CAPES and MEC.

\bibliographystyle{unsrt}
\bibliography{main.bib}

\begin{thebibliography}{10}

\bibitem{dalzell2023quantum}
Alexander~M Dalzell, Sam McArdle, Mario Berta, Przemyslaw Bienias, Chi-Fang
  Chen, Andr{\'a}s Gily{\'e}n, Connor~T Hann, Michael~J Kastoryano, Emil~T
  Khabiboulline, Aleksander Kubica, et~al.
\newblock Quantum algorithms: A survey of applications and end-to-end
  complexities.
\newblock {\em arXiv preprint arXiv:2310.03011}, 2023.

\bibitem{gisin2007quantum}
Nicolas Gisin and Rob Thew.
\newblock Quantum communication.
\newblock {\em Nature photonics}, 1(3):165--171, 2007.

\bibitem{degen2017quantum}
Christian~L Degen, Friedemann Reinhard, and Paola Cappellaro.
\newblock Quantum sensing.
\newblock {\em Reviews of modern physics}, 89(3):035002, 2017.

\bibitem{georgescu2014quantum}
Iulia~M Georgescu, Sahel Ashhab, and Franco Nori.
\newblock Quantum simulation.
\newblock {\em Reviews of Modern Physics}, 86(1):153, 2014.

\bibitem{valivarthi2016quantum}
Raju Valivarthi, Qiang Zhou, Gabriel~H Aguilar, Varun~B Verma, Francesco
  Marsili, Matthew~D Shaw, Sae~Woo Nam, Daniel Oblak, Wolfgang Tittel, et~al.
\newblock Quantum teleportation across a metropolitan fibre network.
\newblock {\em Nature Photonics}, 10(10):676, 2016.

\bibitem{wengerowsky2019entanglement}
S{\"o}ren Wengerowsky, Siddarth~Koduru Joshi, Fabian Steinlechner, Julien~R
  Zichi, Sergiy~M Dobrovolskiy, Ren{\'e} van~der Molen, Johannes~WN Los, Val
  Zwiller, Marijn~AM Versteegh, Alberto Mura, et~al.
\newblock Entanglement distribution over a 96-km-long submarine optical fiber.
\newblock {\em Proceedings of the National Academy of Sciences},
  116(14):6684--6688, 2019.

\bibitem{Bedington2017}
Robert Bedington, Juan~Miguel Arrazola, and Alexander Ling.
\newblock Progress in satellite quantum key distribution.
\newblock {\em npj Quantum Information}, 3(1):30, 2017.

\bibitem{Yin1140}
Juan Yin, Yuan Cao, Yu-Huai Li, Sheng-Kai Liao, Liang Zhang, Ji-Gang Ren,
  Wen-Qi Cai, Wei-Yue Liu, Bo~Li, Hui Dai, Guang-Bing Li, Qi-Ming Lu, Yun-Hong
  Gong, Yu~Xu, Shuang-Lin Li, Feng-Zhi Li, Ya-Yun Yin, Zi-Qing Jiang, Ming Li,
  Jian-Jun Jia, Ge~Ren, Dong He, Yi-Lin Zhou, Xiao-Xiang Zhang, Na~Wang, Xiang
  Chang, Zhen-Cai Zhu, Nai-Le Liu, Yu-Ao Chen, Chao-Yang Lu, Rong Shu,
  Cheng-Zhi Peng, Jian-Yu Wang, and Jian-Wei Pan.
\newblock Satellite-based entanglement distribution over 1200 kilometers.
\newblock {\em Science}, 356(6343):1140--1144, 2017.

\bibitem{PhysRevLett.120.030501}
Sheng-Kai Liao, Wen-Qi Cai, Johannes Handsteiner, Bo~Liu, Juan Yin, Liang
  Zhang, Dominik Rauch, Matthias Fink, Ji-Gang Ren, Wei-Yue Liu, Yang Li,
  Qi~Shen, Yuan Cao, Feng-Zhi Li, Jian-Feng Wang, Yong-Mei Huang, Lei Deng, Tao
  Xi, Lu~Ma, Tai Hu, Li~Li, Nai-Le Liu, Franz Koidl, Peiyuan Wang, Yu-Ao Chen,
  Xiang-Bin Wang, Michael Steindorfer, Georg Kirchner, Chao-Yang Lu, Rong Shu,
  Rupert Ursin, Thomas Scheidl, Cheng-Zhi Peng, Jian-Yu Wang, Anton Zeilinger,
  and Jian-Wei Pan.
\newblock Satellite-relayed intercontinental quantum network.
\newblock {\em Phys. Rev. Lett.}, 120:030501, Jan 2018.

\bibitem{sidhu2021advances}
Jasminder~S Sidhu, Siddarth~K Joshi, Mustafa G{\"u}ndo{\u{g}}an, Thomas
  Brougham, David Lowndes, Luca Mazzarella, Markus Krutzik, Sonali Mohapatra,
  Daniele Dequal, Giuseppe Vallone, et~al.
\newblock Advances in space quantum communications.
\newblock {\em IET Quantum Communication}, 2(4):182--217, 2021.

\bibitem{kimble2008quantum}
H~Jeff Kimble.
\newblock The quantum internet.
\newblock {\em Nature}, 453(7198):1023, 2008.

\bibitem{simon2017towards}
Christoph Simon.
\newblock Towards a global quantum network.
\newblock {\em Nature Photonics}, 11(11):678, 2017.

\bibitem{wehner2018quantum}
Stephanie Wehner, David Elkouss, and Ronald Hanson.
\newblock Quantum internet: A vision for the road ahead.
\newblock {\em Science}, 362(6412):eaam9288, 2018.

\bibitem{caleffi2018quantum}
Marcello Caleffi, Angela~Sara Cacciapuoti, and Giuseppe Bianchi.
\newblock Quantum internet: From communication to distributed computing!
\newblock In {\em Proceedings of the 5th ACM international conference on
  nanoscale computing and communication}, pages 1--4, 2018.

\bibitem{Gisin_2002}
Nicolas Gisin, Grégoire Ribordy, Wolfgang Tittel, and Hugo Zbinden.
\newblock Quantum cryptography.
\newblock {\em Reviews of Modern Physics}, 74(1):145–195, Mar 2002.

\bibitem{buhrman2003distributed}
Harry Buhrman and Hein R{\"o}hrig.
\newblock Distributed quantum computing.
\newblock In {\em International Symposium on Mathematical Foundations of
  Computer Science}, pages 1--20. Springer, 2003.

\bibitem{ho2022entanglement}
Joseph Ho, George Moreno, Samura{\'\i} Brito, Francesco Graffitti,
  Christopher~L Morrison, Ranieri Nery, Alexander Pickston, Massimiliano
  Proietti, Rafael Rabelo, Alessandro Fedrizzi, et~al.
\newblock Entanglement-based quantum communication complexity beyond bell
  nonlocality.
\newblock {\em npj Quantum Information}, 8(1):13, 2022.

\bibitem{hensen2015loophole}
Bas Hensen, Hannes Bernien, Ana{\"\i}s~E Dr{\'e}au, Andreas Reiserer, Norbert
  Kalb, Machiel~S Blok, Just Ruitenberg, Raymond~FL Vermeulen, Raymond~N
  Schouten, Carlos Abell{\'a}n, et~al.
\newblock Loophole-free bell inequality violation using electron spins
  separated by 1.3 kilometres.
\newblock {\em Nature}, 526(7575):682, 2015.

\bibitem{loop1}
Lynden~K. Shalm, Evan Meyer-Scott, Bradley~G. Christensen, Peter Bierhorst,
  Michael~A. Wayne, Martin~J. Stevens, Thomas Gerrits, Scott Glancy, Deny~R.
  Hamel, Michael~S. Allman, Kevin~J. Coakley, Shellee~D. Dyer, Carson Hodge,
  Adriana~E. Lita, Varun~B. Verma, Camilla Lambrocco, Edward Tortorici, Alan~L.
  Migdall, Yanbao Zhang, Daniel~R. Kumor, William~H. Farr, Francesco Marsili,
  Matthew~D. Shaw, Jeffrey~A. Stern, Carlos Abell\'an, Waldimar Amaya, Valerio
  Pruneri, Thomas Jennewein, Morgan~W. Mitchell, Paul~G. Kwiat, Joshua~C.
  Bienfang, Richard~P. Mirin, Emanuel Knill, and Sae~Woo Nam.
\newblock Strong loophole-free test of local realism.
\newblock {\em Phys. Rev. Lett.}, 115:250402, Dec 2015.

\bibitem{loop2}
Marissa Giustina, Marijn A.~M. Versteegh, S\"oren Wengerowsky, Johannes
  Handsteiner, Armin Hochrainer, Kevin Phelan, Fabian Steinlechner, Johannes
  Kofler, Jan-\AA{}ke Larsson, Carlos Abell\'an, Waldimar Amaya, Valerio
  Pruneri, Morgan~W. Mitchell, J\"orn Beyer, Thomas Gerrits, Adriana~E. Lita,
  Lynden~K. Shalm, Sae~Woo Nam, Thomas Scheidl, Rupert Ursin, Bernhard
  Wittmann, and Anton Zeilinger.
\newblock Significant-loophole-free test of bell's theorem with entangled
  photons.
\newblock {\em Phys. Rev. Lett.}, 115:250401, Dec 2015.

\bibitem{pirandola2015advances}
Stefano Pirandola, Jens Eisert, Christian Weedbrook, Akira Furusawa, and
  Samuel~L Braunstein.
\newblock Advances in quantum teleportation.
\newblock {\em Nature photonics}, 9(10):641--652, 2015.

\bibitem{ClockSinc}
P.~K{\'o}m{\'a}r, E.~M. Kessler, M.~Bishof, L.~Jiang, A.~S. Sorensen, J.~Ye,
  and M.~D. Lukin.
\newblock A quantum network of clocks.
\newblock {\em Nature Physics}, 10:582 EP --, Jun 2014.
\newblock Article.

\bibitem{Broadbent_2009}
Anne Broadbent, Joseph Fitzsimons, and Elham Kashefi.
\newblock Universal blind quantum computation.
\newblock {\em 2009 50th Annual IEEE Symposium on Foundations of Computer
  Science}, Oct 2009.

\bibitem{PhysRevA.96.012303}
Joseph~F. Fitzsimons and Elham Kashefi.
\newblock Unconditionally verifiable blind quantum computation.
\newblock {\em Phys. Rev. A}, 96:012303, Jul 2017.

\bibitem{riedel2018europe}
Max~F Riedel, Immanuel Bloch, Thierry Debuisschert, Frank Wilhelm-Mauch,
  Valerio Pruneri, Nikolay~V Vitanov, Stephanie Wehner, and Tommaso Calarco.
\newblock Europe's quantum flagship is taking off.
\newblock {\em Europhysics News}, 49(5-6):30--34, 2018.

\bibitem{awschalom2020long}
David Awschalom.
\newblock From long-distance entanglement to building a nationwide quantum
  internet: Report of the doe quantum internet blueprint workshop.
\newblock Technical report, Brookhaven National Lab.(BNL), Upton, NY (United
  States), 2020.

\bibitem{lu2022micius}
Chao-Yang Lu, Yuan Cao, Cheng-Zhi Peng, and Jian-Wei Pan.
\newblock Micius quantum experiments in space.
\newblock {\em Reviews of Modern Physics}, 94(3):035001, 2022.

\bibitem{ren2017ground}
Ji-Gang Ren, Ping Xu, Hai-Lin Yong, Liang Zhang, Sheng-Kai Liao, Juan Yin,
  Wei-Yue Liu, Wen-Qi Cai, Meng Yang, Li~Li, et~al.
\newblock Ground-to-satellite quantum teleportation.
\newblock {\em Nature}, 549(7670):70--73, 2017.

\bibitem{chen2021integrated}
Yu-Ao Chen, Qiang Zhang, Teng-Yun Chen, Wen-Qi Cai, Sheng-Kai Liao, Jun Zhang,
  Kai Chen, Juan Yin, Ji-Gang Ren, Zhu Chen, et~al.
\newblock An integrated space-to-ground quantum communication network over
  4,600 kilometres.
\newblock {\em Nature}, 589(7841):214--219, 2021.

\bibitem{brito2020statistical}
Samura{\'\i} Brito, Askery Canabarro, Rafael Chaves, and Daniel Cavalcanti.
\newblock Statistical properties of the quantum internet.
\newblock {\em Physical Review Letters}, 124(21):210501, 2020.

\bibitem{brito2021satellite}
Samura{\'\i} Brito, Askery Canabarro, Daniel Cavalcanti, and Rafael Chaves.
\newblock Satellite-based photonic quantum networks are small-world.
\newblock {\em Prx Quantum}, 2(1):010304, 2021.

\bibitem{zhuang2021quantum}
Quntao Zhuang and Bingzhi Zhang.
\newblock Quantum communication capacity transition of complex quantum
  networks.
\newblock {\em Physical Review A}, 104(2):022608, 2021.

\bibitem{azuma2021tools}
Koji Azuma, Stefan B{\"a}uml, Tim Coopmans, David Elkouss, and Boxi Li.
\newblock Tools for quantum network design.
\newblock {\em AVS Quantum Science}, 3(1), 2021.

\bibitem{harney2022analytical}
Cillian Harney and Stefano Pirandola.
\newblock Analytical methods for high-rate global quantum networks.
\newblock {\em PRX Quantum}, 3(1):010349, 2022.

\bibitem{bugalho2023distributing}
Lu{\'\i}s Bugalho, Bruno~C Coutinho, Francisco~A Monteiro, and Yasser Omar.
\newblock Distributing multipartite entanglement over noisy quantum networks.
\newblock {\em quantum}, 7:920, 2023.

\bibitem{wei2022towards}
Shi-Hai Wei, Bo~Jing, Xue-Ying Zhang, Jin-Yu Liao, Chen-Zhi Yuan, Bo-Yu Fan,
  Chen Lyu, Dian-Li Zhou, You Wang, Guang-Wei Deng, et~al.
\newblock Towards real-world quantum networks: a review.
\newblock {\em Laser \& Photonics Reviews}, 16(3):2100219, 2022.

\bibitem{nokkala2020probing}
Johannes Nokkala, Jyrki Piilo, and Ginestra Bianconi.
\newblock Probing the spectral dimension of quantum network geometries.
\newblock {\em Journal of Physics: Complexity}, 2(1):015001, 2020.

\bibitem{zhang2021quantum}
Bingzhi Zhang and Quntao Zhuang.
\newblock Quantum internet under random breakdowns and intentional attacks.
\newblock {\em Quantum Science and Technology}, 6(4):045007, 2021.

\bibitem{albert2002statistical}
R{\'e}ka Albert and Albert-L{\'a}szl{\'o} Barab{\'a}si.
\newblock Statistical mechanics of complex networks.
\newblock {\em Reviews of modern physics}, 74(1):47, 2002.

\bibitem{barabasi2016network}
Albert-L{\'a}szl{\'o} Barab{\'a}si et~al.
\newblock {\em Network science}.
\newblock Cambridge university press, 2016.

\bibitem{nokkala2023complex}
Johannes Nokkala, Jyrki Piilo, and Ginestra Bianconi.
\newblock Complex quantum networks: a topical review.
\newblock {\em arXiv preprint arXiv:2311.16265}, 2023.

\bibitem{zukowski1993event}
Marek {\.Z}ukowski, Anton Zeilinger, Michael~A Horne, and Artur~K Ekert.
\newblock Event-ready-detectors bell experiment via entanglement swapping.
\newblock {\em Physical Review Letters}, 71(26):4287, 1993.

\bibitem{pan1998experimental}
Jian-Wei Pan, Dik Bouwmeester, Harald Weinfurter, and Anton Zeilinger.
\newblock Experimental entanglement swapping: entangling photons that never
  interacted.
\newblock {\em Physical Review Letters}, 80(18):3891, 1998.

\bibitem{briegel1998quantum}
H-J Briegel, Wolfgang D{\"u}r, Juan~I Cirac, and Peter Zoller.
\newblock Quantum repeaters: the role of imperfect local operations in quantum
  communication.
\newblock {\em Physical Review Letters}, 81(26):5932, 1998.

\bibitem{sangouard2011quantum}
Nicolas Sangouard, Christoph Simon, Hugues De~Riedmatten, and Nicolas Gisin.
\newblock Quantum repeaters based on atomic ensembles and linear optics.
\newblock {\em Reviews of Modern Physics}, 83(1):33, 2011.

\bibitem{ruihong2019research}
Qiao Ruihong and Meng Ying.
\newblock Research progress of quantum repeaters.
\newblock In {\em Journal of Physics: Conference Series}, volume 1237, page
  052032. IOP Publishing, 2019.

\bibitem{Azuma22}
David Elkouss Paul Hilaire Liang Jiang Hoi-Kwong~Lo Koji~Azuma, Sophia
  E.~Economou and Ilan Tzitrin.
\newblock Quantum algorithms: A survey of applications and end-to-end
  complexities.
\newblock {\em arXiv preprint arXiv:2212.10820}, 2022.

\bibitem{ipea}
Rafael H.~M. Pereira and Caio~Nogueira Goncalves.
\newblock {\em geobr: Download Official Spatial Data Sets of Brazil}, 2023.
\newblock R package version 1.8.1.

\bibitem{bernardes2011rate}
Nadja~K Bernardes, Ludmi{\l}a Praxmeyer, and Peter van Loock.
\newblock Rate analysis for a hybrid quantum repeater.
\newblock {\em Physical Review A}, 83(1):012323, 2011.

\bibitem{lakhina2002geographic}
Anukool Lakhina, John~W Byers, Mark Crovella, and Ibrahim Matta.
\newblock On the geographic location of internet resources.
\newblock In {\em Proceedings of the 2nd ACM SIGCOMM Workshop on Internet
  measurment}, pages 249--250, 2002.

\bibitem{waxman1988routing}
Bernard~M Waxman.
\newblock Routing of multipoint connections.
\newblock {\em IEEE journal on selected areas in communications},
  6(9):1617--1622, 1988.

\bibitem{durairajan2015intertubes}
Ramakrishnan Durairajan, Paul Barford, Joel Sommers, and Walter Willinger.
\newblock Intertubes: A study of the us long-haul fiber-optic infrastructure.
\newblock In {\em Proceedings of the 2015 ACM Conference on Special Interest
  Group on Data Communication}, pages 565--578, 2015.

\bibitem{gisin2015far}
Nicolas Gisin.
\newblock How far can one send a photon?
\newblock {\em Frontiers of Physics}, 10:1--8, 2015.

\bibitem{qiu2010financial}
Tian Qiu, Bo~Zheng, and Guang Chen.
\newblock Financial networks with static and dynamic thresholds.
\newblock {\em New Journal of Physics}, 12(4):043057, 2010.

\bibitem{valente2004two}
Andr{\'e}~XCN Valente, Abhijit Sarkar, and Howard~A Stone.
\newblock Two-peak and three-peak optimal complex networks.
\newblock {\em Physical Review Letters}, 92(11):118702, 2004.

\bibitem{bennett1993teleporting}
Charles~H Bennett, Gilles Brassard, Claude Cr{\'e}peau, Richard Jozsa, Asher
  Peres, and William~K Wootters.
\newblock Teleporting an unknown quantum state via dual classical and
  einstein-podolsky-rosen channels.
\newblock {\em Physical review letters}, 70(13):1895, 1993.

\bibitem{buhrman2001quantum}
Harry Buhrman, Richard Cleve, and Wim Van~Dam.
\newblock Quantum entanglement and communication complexity.
\newblock {\em SIAM Journal on Computing}, 30(6):1829--1841, 2001.

\bibitem{scarani2009security}
Valerio Scarani, Helle Bechmann-Pasquinucci, Nicolas~J Cerf, Miloslav
  Du{\v{s}}ek, Norbert L{\"u}tkenhaus, and Momtchil Peev.
\newblock The security of practical quantum key distribution.
\newblock {\em Reviews of modern physics}, 81(3):1301, 2009.

\bibitem{pan2001entanglement}
Jian-Wei Pan, Christoph Simon, {\v{C}}aslav Brukner, and Anton Zeilinger.
\newblock Entanglement purification for quantum communication.
\newblock {\em Nature}, 410(6832):1067--1070, 2001.

\bibitem{PhysRevLett.112.250501}
Sreraman Muralidharan, Jungsang Kim, Norbert L\"utkenhaus, Mikhail~D. Lukin,
  and Liang Jiang.
\newblock Ultrafast and fault-tolerant quantum communication across long
  distances.
\newblock {\em Phys. Rev. Lett.}, 112:250501, Jun 2014.

\bibitem{PhysRevLett.130.213601}
V.~Krutyanskiy, M.~Canteri, M.~Meraner, J.~Bate, V.~Krcmarsky, J.~Schupp,
  N.~Sangouard, and B.~P. Lanyon.
\newblock Telecom-wavelength quantum repeater node based on a trapped-ion
  processor.
\newblock {\em Phys. Rev. Lett.}, 130:213601, May 2023.

\bibitem{PhysRevA.87.052315}
Silvestre Abruzzo, Sylvia Bratzik, Nadja~K. Bernardes, Hermann Kampermann,
  Peter van Loock, and Dagmar Bru\ss{}.
\newblock Quantum repeaters and quantum key distribution: Analysis of
  secret-key rates.
\newblock {\em Phys. Rev. A}, 87:052315, May 2013.

\bibitem{PhysRevA.100.032322}
E.~Shchukin, F.~Schmidt, and P.~van Loock.
\newblock Waiting time in quantum repeaters with probabilistic entanglement
  swapping.
\newblock {\em Phys. Rev. A}, 100:032322, Sep 2019.

\bibitem{https://doi.org/10.1002/qute.201900141}
Peter van Loock, Wolfgang Alt, Christoph Becher, Oliver Benson, Holger Boche,
  Christian Deppe, Jürgen Eschner, Sven Höfling, Dieter Meschede, Peter
  Michler, Frank Schmidt, and Harald Weinfurter.
\newblock Extending quantum links: Modules for fiber- and memory-based quantum
  repeaters.
\newblock {\em Advanced Quantum Technologies}, 3(11):1900141, 2020.

\bibitem{liorni2021quantum}
Carlo Liorni, Hermann Kampermann, and Dagmar Bru{\ss}.
\newblock Quantum repeaters in space.
\newblock {\em New Journal of Physics}, 23(5):053021, 2021.

\bibitem{PhysRevA.105.012608}
Tim Coopmans, Sebastiaan Brand, and David Elkouss.
\newblock Improved analytical bounds on delivery times of long-distance
  entanglement.
\newblock {\em Phys. Rev. A}, 105:012608, Jan 2022.

\bibitem{PhysRevA.107.012609}
Guus Avis, Filip Rozp\ifmmode~\mbox{\k{e}}\else \k{e}\fi{}dek, and Stephanie
  Wehner.
\newblock Analysis of multipartite entanglement distribution using a central
  quantum-network node.
\newblock {\em Phys. Rev. A}, 107:012609, Jan 2023.

\bibitem{dijkstra1959note}
Edsger~W Dijkstra et~al.
\newblock A note on two problems in connexion with graphs.
\newblock {\em Numerische mathematik}, 1(1):269--271, 1959.

\bibitem{Azuma15}
Tamaki~K. Azuma, K. and HK. Lo.
\newblock All-photonic quantum repeaters.
\newblock {\em Nat. Commun.}, 6:6787, 2015.

\bibitem{PhysRevA.92.052324}
Seung-Woo Lee, Kimin Park, Timothy~C. Ralph, and Hyunseok Jeong.
\newblock Nearly deterministic bell measurement with multiphoton entanglement
  for efficient quantum-information processing.
\newblock {\em Phys. Rev. A}, 92:052324, Nov 2015.

\bibitem{doi:10.1126/science.aan0070}
N.~Kalb, A.~A. Reiserer, P.~C. Humphreys, J.~J.~W. Bakermans, S.~J. Kamerling,
  N.~H. Nickerson, S.~C. Benjamin, D.~J. Twitchen, M.~Markham, and R.~Hanson.
\newblock Entanglement distillation between solid-state quantum network nodes.
\newblock {\em Science}, 356(6341):928--932, 2017.

\bibitem{doi:10.1126/science.abg1919}
M.~Pompili, S.~L.~N. Hermans, S.~Baier, H.~K.~C. Beukers, P.~C. Humphreys,
  R.~N. Schouten, R.~F.~L. Vermeulen, M.~J. Tiggelman, L.~dos Santos~Martins,
  B.~Dirkse, S.~Wehner, and R.~Hanson.
\newblock Realization of a multinode quantum network of remote solid-state
  qubits.
\newblock {\em Science}, 372(6539):259--264, 2021.

\bibitem{PhysRevX.10.021071}
Johannes Borregaard, Hannes Pichler, Tim Schr\"oder, Mikhail~D. Lukin, Peter
  Lodahl, and Anders~S. S\o{}rensen.
\newblock One-way quantum repeater based on near-deterministic photon-emitter
  interfaces.
\newblock {\em Phys. Rev. X}, 10:021071, Jun 2020.

\bibitem{PhysRevLett.98.060502}
O.~A. Collins, S.~D. Jenkins, A.~Kuzmich, and T.~A.~B. Kennedy.
\newblock Multiplexed memory-insensitive quantum repeaters.
\newblock {\em Phys. Rev. Lett.}, 98:060502, Feb 2007.

\end{thebibliography}

\end{document}